\documentclass[aps,prb,twocolumn,amsmath,amssymb,superscriptaddress]{revtex4-1}

\usepackage{graphicx}
\usepackage{dcolumn}
\usepackage{bm}
\usepackage{amsmath}
\usepackage{color}
\usepackage[normalem]{ulem}

\begin{document}

\title{
Microscopic origins of the anomalous melting behaviour of high-pressure sodium
}

\author{Hagai Eshet}
\email{hagai.eshet@gmail.com}
\affiliation{
Department of Chemistry and Applied Biosciences, ETH Z\"urich, USI Campus, via G. Buffi 13, 6900 Lugano, Switzerland
}
\author{Rustam Z. Khaliullin}
\email{rustam@khaliullin.com}
\affiliation{
Department of Chemistry and Applied Biosciences, ETH Z\"urich, USI Campus, via G. Buffi 13, 6900 Lugano, Switzerland
}
\affiliation{
Physical Chemistry Institute, University of Z\"urich, Winterthurerstrasse 190, 8057 Z\"urich, Switzerland
}
\author{Thomas D. K\"uhne}
\affiliation{
Institute of Physical Chemistry, Johannes Gutenberg University Mainz, D-55128 Mainz, Germany
}
\affiliation{
Center for Computational Sciences, Johannes Gutenberg University Mainz, D-55128 Mainz, Germany
}
\author{J\"org Behler} \affiliation{
Lehrstuhl f\"ur Theoretische Chemie, Ruhr-Universit\"at Bochum, D-44780 Bochum, Germany
}
\author{Michele Parrinello}
\affiliation{
Department of Chemistry and Applied Biosciences, ETH Z\"urich, USI Campus, via G. Buffi 13, 6900 Lugano, Switzerland
}

\date{\today}

\maketitle

\textbf{Recent experiments have shown that sodium, a prototype simple metal at ambient conditions, exhibits unexpected complexity under high pressure~\cite{a:gregorprl,a:mcmahon,a:gregor,a:na-oganov}. One of the most puzzling phenomena in the behaviour of dense sodium is the pressure-induced drop in its melting temperature, which extends from 1000~K at $\sim$30~GPa to as low as room temperature at $\sim$120~GPa~\cite{a:gregorprl}. Despite significant theoretical effort to understand the anomalous melting~\cite{a:hernandez,a:raty,a:pinsook,a:yamane,a:Na-martinez-canales,a:lepeshkin} its origins have remained unclear. In this work, we reconstruct the sodium phase diagram using an \emph{ab-initio}-quality neural-network potential~\cite{a:behler,a:khalNa1}. We demonstrate that the reentrant behaviour results from the screening of interionic interactions by conduction electrons, which at high pressure induces a softening in the short-range repulsion. It is expected that such an effect plays an important role in governing the behaviour of a wide range of metals and alloys.}

\begin{figure}
\includegraphics*[width=8cm]{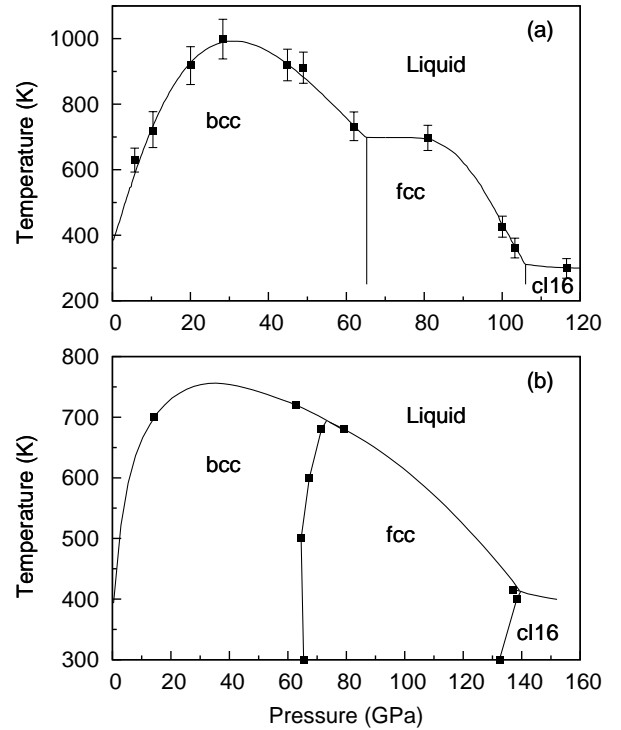}\\ 
\caption{\label{fig:phd} Sodium phase diagram. a. The squares represent experimentally measured melting points. The lines show tentative phase boundaries~\cite{a:gregorprl}.
b. Theoretical results based on the NN potential. The squares represent coexistence points computed with thermodynamic integration. The lines show coexistence curves traced by integrating the Clausius--Clapeyron equation (see Methods).}
\end{figure}

At ambient conditions, sodium crystallizes in the body-centered cubic (bcc) structure, which transforms to the face-centered cubic (fcc) phase upon compression to 65~GPa~\cite{a:hanfland}. At 105~GPa, the fcc phase undergoes a transformation to the cI16 phase -- a distorted variant of the bcc structure~\cite{a:mcmahon}. Further compression results in the formation of a large variety of complex phases~\cite{a:gregor,a:gregor,a:na-oganov}, the existence of which is quite intriguing.
However, the focus of this work is one of the most striking features of the sodium phase diagram - the anomalous melting curve. Diffraction measurements have revealed a maximum in melting temperature around 1000~K at $\sim$30~GPa followed by a pressure-induced drop, which extends to nearly room temperature at $\sim$120~GPa~\cite{a:gregorprl} and spans the regions of stability of three solid phases (Fig.~\ref{fig:phd}a). This anomaly implies that in this range the liquid is denser than the solid.

Although previous \emph{ab initio} studies~\cite{a:hernandez,a:raty,a:pinsook,a:yamane,a:Na-martinez-canales,a:lepeshkin} have reproduced the reentrant melting behaviour of sodium the physical origin of this anomaly still remains unclear. The hypothesis of structural and electronic transitions in liquid sodium proposed in one of the studies~\cite{a:raty} has not been confirmed by other authors~\cite{a:hernandez}. The discrepancy between these simulation results has been attributed to a different treatment of the semicore states~\cite{a:yamane}.

It is important to point out that due to the high computational cost of \emph{ab initio} MD all simulations so far reported~\cite{a:hernandez,a:raty,a:pinsook,a:yamane} have used small simulation cells and estimated melting temperatures ($T_m$) by heating the solid phase until it melts. In such simulations, the absence of nucleation centers and fast heating schedules delay melting relative to the thermodynamic coexistence point and, therefore, only an upper bound on $T_m$ can be obtained. A proper reconstruction of the melting curve requires comparing the free energies of the liquid and solid phases and involves long ($\sim$10~ns) MD simulations on large systems ($\sim 10^3$ atoms). Performing such simulations with direct \emph{ab initio} methods is computationally too expensive at present. Although several authors~\cite{a:Na-martinez-canales,a:lepeshkin} have shown that simplified models can also reproduce the anomalous melting behavior the relation of these models to the actual physical interactions in sodium has remained unclear.

In a previous work~\cite{a:khalNa1}, we have shown that a neural-network (NN) potential for sodium based on well-converged electronic structure calculations, retains the accuracy of \emph{ab initio} simulations at a highly reduced computational cost. The ability of the NN potential to reproduce numerous \emph{ab initio} and experimental properties of the solid and the liquid phases in the pressure range up to 140~GPa has been shown in Ref.~\onlinecite{a:khalNa1}. Additionally, Supplementary Fig.~1  demonstrates high accuracy of the NN potential in reproducing the \emph{ab initio} energies for liquid sodium structures along MD trajectories in the relevant P--T range.

Here we utilize this NN potential to construct the sodium phase diagram. The phase diagram shown in Fig.~\ref{fig:phd}b is calculated using thermodynamic integration and by tracing the coexistence curve using the Clausius-Clapeyron equation (see Methods). The NN potential captures the experimentally observed sequence of the solid-state phase transitions bcc$\rightarrow$fcc$\rightarrow$cI16 and the regions of stability of each phase. The shape of the melting curve with the maximum around 30~GPa is also correctly reproduced. The comparison of NN and DFT transition pressures at zero temperature (Fig.~6 in Ref.~\onlinecite{a:khalNa1}) as well as the ``heat-until-it-melts`` melting curves obtained from NN and \emph{ab initio} simulations (Supplementary Fig.~2) show that the quantitative errors in coexistence lines can be attributed to inaccuracies of the density functional and not to the errors of the NN fitting. Supplementary Fig.~2 also demonstrates that $T_m$ obtained with a non-equilibrium approach is overestimated compared to the proper thermodynamic calculation. To the best of our knowledge this is the first phase diagram calculated from a high-quality \emph{ab initio} method. A careful analysis of structural and electronic properties of liquid sodium shows no evidence of liquid-liquid phase transitions proposed earlier~\cite{a:raty} (see Supplementary Information).

While the NN simulations reproduce adequately the sodium melting curve the physical origins of its reentrant behaviour are difficult to extract. To this effect, we constructed a density dependent pair potential based on the jellium model for the conducting electrons. According to this model the ionic structure of a metal can be replaced to a first approximation, by a uniform positively-charged background while conduction electrons can be treated as a uniform electron gas. The granularity of sodium is then introduced at the level of two-body interactions by immersing two Na atoms in such a jellium and calculating the energy as a function of the interatomic distance (see Methods). The effect of the pressure is taken into account by varying the density of the electron gas $\rho_e$ according to the NN equation of state. By doing so, we retrace a time-honoured idea used to construct effective potentials describing screening effects in metals by linear response theory~\cite{b:ashcroft}. Here we use a non-linear approach since it is expected that non-linear screening effects become important as the ions are brought closer to each other by the applied pressure.

We found that the two-body effective potential obtained from the jellium model is well reproduced by a sum of the repulsive Yukawa potential and the oscillatory Friedel term (Supplementary Fig.~3)~\cite{a:boon,a:paskin-rahman}:
\begin{equation}
\label{eq:jellium}
\phi (r) =  \frac{A \exp (- k_0 r)}{r}  + \frac{B \cos [2 k_F (r - r_0)]}{r^m},
\end{equation}
where $k_F$ is the Fermi wavenumber and $A$, $B$, $k_0$, $r_0$ and $m$ are density dependent fitting parameters (see Methods). This analytical form of the potential is inspired by a linear screening theory~\cite{b:ashcroft} and reflects the nature of physical interactions in metallic sodium, in which the direct Na--Na repulsion is screened by the free-electron gas. In particular, the oscillatory term has its origin in the sharpness of the Fermi surface. 

\begin{figure}
\includegraphics*[width=8cm]{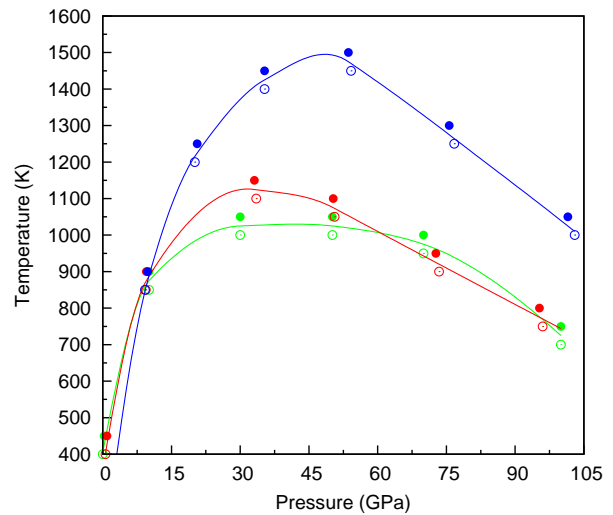}\\ 
\caption{\label{fig:huim} The "heat-until-it-melts" curves obtained with the NN potential (green), effective pair potential based on the jellium model (red) and repulsive wall of the effective pair potential (blue). The points below 90~GPa are obtained by melting the bcc phase, the points above 90~GPa by melting the fcc phase.
}
\end{figure}

MD simulations based on the effective pair potential reproduce semi-quantitatively the maximum in the melting curve (Fig.~\ref{fig:huim}). The potential is also remarkably accurate in predicting the radial distribution functions (RDFs) of the liquid for pressures up to 100~GPa (Supplementary Fig.~4). We thus believe that this effective density dependent two-body potential is able to capture the physics of the problem and can be used to shed light on the origin of the anomalous melting behaviour. It is worth mentioning that the validity of this potential is limited to the region of pressures below about 120 GPa because beyond this region the electron are promoted to the $d$-states and the simple free-electron two-body model is no longer appropriate.

We have examined the effect of two different parts of the effective potential on the melting curve. First, we considered the Yukawa part of the potential and, second, the repulsive wall of $\phi (r)$. The latter was constructed by truncating the total potential at its first minimum and shifting it so that both the energy and force are zero at the truncation point~\cite{a:wca}. We found that the Yukawa potential cannot reproduce the reentrant melting behaviour. On the contrary, the melting curve obtained with the repulsive wall exhibits a maximum similar to that obtained with the full potential (Fig.~\ref{fig:huim}). This observation suggests that it is the short range repulsive part of the potential that provides the origin of the anomalous melting (see Supplementary information for a detailed pressure decomposition analysis that confirms this conclusion).

\begin{figure}
\includegraphics*[width=8cm]{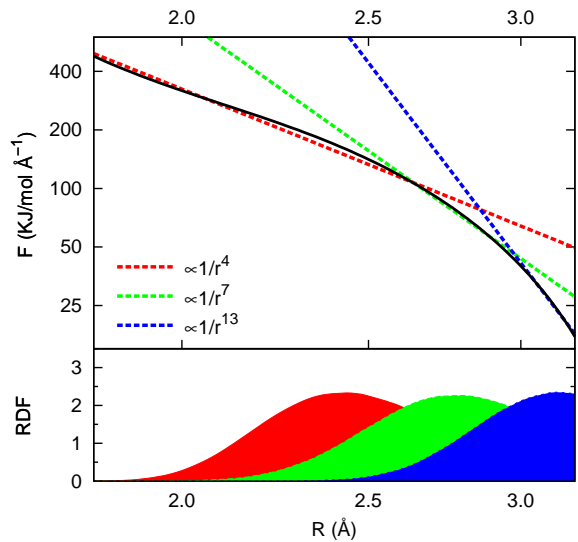}\\ 
\caption{\label{fig:switch} Upper panel: interatomic forces in the region of the repulsive wall of the effective pair potential (black solid line). Lower panel: RDF of liquid sodium along the melting curve at $\sim$10~GPa (blue), $\sim$30~GPa (green) and $\sim$90~GPa (red).}
\end{figure}

The log-log plot of the interatomic force in the region of the repulsive potential wall shown in Fig.~\ref{fig:switch}) supports this point of view. It is seen that there is a large change of the slope as the potential changes from $1/r^{12}$ to a much softer $1/r^3$ behaviour (i.e. the slope changes from $1/r^{13}$ to $1/r^4$ in Fig.~\ref{fig:switch}). The lower panel of the same figure shows the RDFs obtained at different pressures along the melting line. It demonstrates that at low pressure only the steep part of the repulsive wall is sampled, whereas at higher pressure the softer short-range part of the potential starts influencing the atomic interactions. Since softening the potential lowers the melting temperature~\cite{a:inversep,b:phasediagrams} we argue that the softening of the repulsive interactions in sodium at high pressure is the origin of the observed anomaly.

Our results fit well into the existing theories of metals that predict increase of screening with density and possible softening of pair interactions~\cite{a:hafner2,a:hafner1}. They are also in agreement with the fact that certain softening of pair interactions can result in anomalous melting~\cite{a:prestipino}.

\begin{figure}
\includegraphics*[width=8cm]{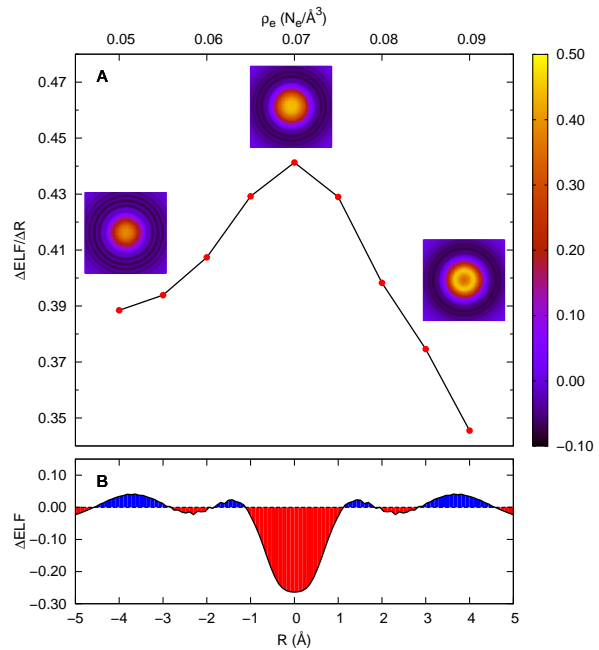}\\ 
\caption{\label{fig:rho} a. Pressure dependence of the derivative of the ELF at the center of the Na--Na bond wrt interatomic distance taken at the distance of the nearest neighbour in the bcc lattice. For comparison, the maximum of $T_m$ is around $\rho_e\approx 0.0615 \AA^{-3}$. The insets show 2D plots of the same derivative in the plane perpendicular to the Na--Na bond for $\rho_e$ 0.05~$\AA^{-3}$ ($\sim$20~GPa), 0.07~$\AA^{-3}$ ($\sim$45~GPa) and 0.09~$\AA^{-3}$ ($\sim$90~GPa). b. Difference between the high-pressure and low-pressure values of ELF in the direction perpendicular to the Na--Na bond, $\Delta$ELF$\equiv$ELF(90~GPa)$-$ELF(25~GPa).}
\end{figure}

In order to understand the electronic origin of the softening effect we examined the behaviour of the electron localization function (ELF), which is widely used to analyse the nature of chemical bonds (see Methods)~\cite{a:elfbecke,a:elfsavin}. Fig.~\ref{fig:rho}a shows that the behaviour of the derivative of the ELF with respect to the interatomic distance, which can be taken as a measure of the softness of the bond, closely follows the behaviour of the melting curve dramatically decreasing at high pressure. It is instructive to compare the ELFs at low and high pressures. Fig.~\ref{fig:rho}b shows that electrons localized in the Na--Na bond move to the outer regions as pressure increases. The tendency of electrons to move from the bonding regions to the interstitial regions at high pressure has already been noted~\cite{a:interstitial1,a:interstitial2}.

In conclusion, the insight into the electronic and structural properties of sodium obtained in this work offer a new consistent explanation of the anomalous melting behaviour of Na. We demonstrate that the observed dramatic drop in the melting temperature can be attributed to the density dependence of the conduction electrons screening, which induces a softening of the interatomic potential at short range. These findings have immediate implications for explaining behaviour of other metals and alloys.

\textbf{Methods.} \emph{NN potential.} The NN potential~\cite{a:behler} was created to reproduce the Perdew-Burke-Ernzerhof (PBE) density functional energies of liquid sodium and bcc, fcc, cI16 crystal phases in the P--T region up to 140~GPa and 1200~K, as described in our previous paper~\cite{a:khalNa1}. An ultrasoft pseudopotential with the $2s$ and $2p$ semicore electrons included explicitly as the valence states was used to describe the energetics of the high-pressure structures.  A large plane-wave cutoff of 100~Ry and a dense mesh of $k$-points were used for all structures to ensure convergence of the total energy up to 1.5~meV/atom. The NN fitting procedure only introduces a small error (RMSE of the independent test sets is 0.91~meV/atom) in addition to the numerical (convergence) error of the DFT calculations. Thus, the NN-potential closely reproduces the \emph{ab initio} potential energy surface and describes all properties and processes in high-pressure high-temperature sodium with an accuracy comparable with that of the PBE functional. The accuracy of the NN potential for numerous structural and dynamical properties of all solid and liquid phases in the relevant P--T range has been demonstrated in our previous paper~\cite{a:khalNa1}. Supplementary Fig.~2 shows that, in addition to these properties, the NN potential reproduces a ``heat-until-it-melts`` melting curve obtained from \emph{ab initio} simulations~\cite{a:yamane}. 

\emph{Phase diagram.} The coexistence lines were determined by locating points of equal chemical potential for each pair of phases in the P--T plane. This was done in three steps (see Ref.~\onlinecite{a:khalcoex}). First, we calculated the Helmholtz free energy of each pair of phases by thermodynamic integration using the Einstein crystal and the Lennard-Jones potential as the reference systems~\cite{b:frenkel2001}. In the next step, the chemical potentials were evaluated by integrating the free energy as a function of the density. Finally, the coexistence lines were traced by integrating the Clausius-Clapeyron equation using the predictor-corrector scheme of Kofke~\cite{a:kofke}.

In the first step, NN-driven MD simulations were performed for systems containing several hundred atoms (1024 atoms -- bcc, cI16, liquid; 868 atoms -- fcc). The temperature was controlled using a coloured-noise Langevin thermostat that was tuned to provide the optimum sampling efficiency over all relevant vibrational modes~\cite{a:ceriotti}. The time step was set to 2~fs. The integral in the first step was evaluated numerically by the Gauss-Legendre quadrature with 10 points. At each value of the coupling parameter $\lambda$, the average value of the integrand and its statistical error were obtained from a 100~ps trajectory. In the second step, the state points along the isotherms were obtained from NPT simulations (2000 atoms -- bcc and cI16, 2048 atoms -- fcc and liquid) governed by the Nos\'e--Hoover equations of motion with Langevin noise on the particle and cell velocities~\cite{a:ceriotti}. Averaging over a 200~ps trajectory was performed for each state point. The predictor-corrector algorithm was iterated until the melting temperature had converged to less than 2~K, which required 2--4 iterations of 100~ps each.

The total simulation time required to model the phase diagram amounts to $\sim$50~ns clearly demonstrating the computational advantage of the NN approach in comparison with the direct \textit{ab initio} simulation.

\emph{Jellium model.} Two sodium atoms were placed in a periodically replicated 17.2~\AA\ cubic cell containing free electrons with a compensating positive uniform background charge. To reproduce the compression effect, the number of free electrons was varied so that their density is in agreement with that of free $3s$ electrons in compressed sodium. To determine the pair potential, the ground-state energy of the system was computed for a series of fixed interatomic distances with the PBE density functional and an ultrasoft semicore pseudopotential. A 30~Ry plane-wave cutoff, 1000~K smearing temperature and a $4\times 4\times 4$ Monkhorst--Pack $k$-point mesh were used to converge the potential curves to 1~meV. The Quantum-Espresso package was used to perform all electronic structure calculations. The calculated DFT energies were fitted with the analytical form in Eq.~\ref{eq:jellium}. The fitting parameters are shown in Supplementary Table~1.

RDFs obtained with the effective pair potential (Supplementary Fig.~4) perfectly reproduce those from the NN-driven simulations. Above $\sim$100~GPa (i.e. in the region of the cI16 phase) the pair potential fails to reproduce the more accurate NN simulations because of the increasing contribution of many-body interactions.

\emph{Electron localization function.} The ELF is calculated following the definition given in Ref.~\onlinecite{a:elfbecke}
\begin{equation}
\label{eq:elf}
\text{ELF} (r) =  \frac{1}{1+(D(r)/D_h(r))^2},
\end{equation}
where
\begin{equation}
\label{eq:d}
D(r) =  \frac{1}{2} \sum_i \vert \nabla \psi_i(r) \vert^2 - \frac{1}{8} \frac{\vert \nabla \rho(r) \vert^2}{\rho(r)}
\end{equation}
is a measure of the electron localization and
\begin{equation}
\label{eq:d0}
D_h(r) =  \frac{3}{10} (3 \pi^2)^{2/3} \rho(r)^{5/3}
\end{equation}
is the reference uniform electron gas value.

High ELF values show that at the examined position the electrons are more localized than in a uniform electron gas of the same density, whereas ELF$(r)=0.5$ indicates that the localization effect is the same as in the uniform electron gas.

\textbf{Acknowledgments.} This work was supported by the European Research Council (ERC-2009-AdG-247075). J.B. is grateful for financial support from the FCI and the DFG. T.D.K. acknowledges support by the Graduate School of Excellence MAINZ. Our thanks are also due to the Swiss National Supercomputing Centre (CSCS) and High Performance Computing Group of ETH Z\"urich for computer time.

\bibliography{Na}

\end{document}